\documentclass[letterpaper]{article} 
\usepackage{aaai2026}  
\usepackage{times}  
\usepackage{helvet}  
\usepackage{courier}  
\usepackage[hyphens]{url}  
\usepackage{graphicx} 
\usepackage{natbib}  
\usepackage{caption} 
\usepackage{booktabs} 
\usepackage{xcolor}
\usepackage[most]{tcolorbox}
\usepackage{longtable}

\usepackage{algorithm}
\usepackage{algorithmic}

\usepackage{newfloat}
\usepackage{listings}
\DeclareCaptionStyle{ruled}{labelfont=normalfont,labelsep=colon,strut=off} 
\floatstyle{ruled}
\newfloat{listing}{tb}{lst}{}
\floatname{listing}{Listing}
\title{The Meme Is the Message: Generative Memesis and AI Visuals in the 2024 USA Presidential Elections}
\author{
    Ho-Chun Herbert Chang\textsuperscript{\rm 1}, 
    Yung-Chun Chen\textsuperscript{\rm 1}, 
    Benjamin Shaman\textsuperscript{\rm 1},
    Mingyue Zha\textsuperscript{\rm 1}, 
    Sean Noh\textsuperscript{\rm 1},  \\
    Chiyu Wei\textsuperscript{\rm 2},
    Tracy Weener\textsuperscript{\rm 1}, 
    Maya Magee\textsuperscript{\rm 1}
}
\affiliations{
    \textsuperscript{\rm 1} Program in Quantitative Social Science, Dartmouth College, Hanover, NH 03755 USA \\
    \textsuperscript{\rm 2} Department of Mathematics, Dartmouth College, Hanover, NH 03755 USA
}

\usepackage{bibentry}

\begin{document}

\maketitle

\begin{abstract}
Visual content on social media has become increasingly influential in shaping political discourse and civic engagement, but it also limits participation due to the increased cost of multimedia production. In tandem, the growth of generative AI provides novel ways for citizens to participate in politics by lowering these costs. Drawing on a dataset of 239,526 Instagram images, we analyze the effects of synthetic images during the 2024 United States presidential election, using a multimodal workflow combining computer vision, large language models, and facial affect analysis. Results show that meme format is a stronger predictor of engagement than AI-generated content alone. However, AI-generated memes yield a significant interaction effect, suggesting synergistic increases in engagement when synthetic imagery is integrated with memes through human curation. We also characterize how users curate images. Partisans use AI in different ways: Democrat-leaning users tend to use it for in-group support, whereas Republican-leaning users more often employ it for out-group attacks. Users generally select happier synthetic faces compared to real photographs. We define generative memesis as a mode of communication in which memes are no longer shared person-to-person, but mediated by AI through customized visuals. We discuss how generative AI may empower civic participation, the bifurcation of content production and curation, and its implications for in the history of novel technologies and participatory culture.
\end{abstract}


\section{Introduction}

Recent advancement of large-language models and generative AI have spurred similar excitement and concern about their impact on information environments \citep{Argyle2023,Coeckelbergh2019,Delacroix2025,chang2025language}. Broadly, generative AI refers to a class of artificial intelligence models capable of producing novel content—such as text, images, video, or audio—by learning patterns from large datasets and generating outputs that resemble the data they were trained on \citep{EpsteinHertzmann2023}. The 2024 United States election was the first U.S.\ presidential election that intersected these technologies, raising concerns about the possible impact of generative AI \citep{Lee2024TIME}. Even prior to the U.S.\ elections, Slovakia saw the use of deepfaked voices to spread misinformation \citep{DeNadal2024}. Images and videos were also a crucial vehicle for misinformation in other elections around the world and motivated work on media literacy mechanisms for suppressing visual misinformation \citep{Qian2023,GarimellaEckles2020,ChangWangFang2024}. Considerable recent research has also been devoted to understanding mechanistically the role of generative AI on (mis)information processing \citep{ShinKoerberLim2024,Ou2024}.

Visuals have a long history in civic engagement and political mobilization, ranging from iconic recruitment posters in World War I. Compared to the single modality of text, visuals command more attention \citep{Galloway2017}, elicit empathy toward activists \citep{BrownMourao2021}, and evoke stronger emotional responses \citep{Iyer2014}. In the present, visual social media platforms like Instagram and TikTok have generated new forms of engagement, ranging from activism \citep{Chapman2016}, social movements \citep{ChangRichardsonFerrara2022}, and civic participation \citep{MoffettRice2024}, that were once typical of text-based platforms like Twitter. Meme creation, for instance, spurred creative, multimodal political engagement from everyday users \citep{Shifman2013a,McLoughlinSouthern2021}. Additionally, multimedia has increased barriers to participation given the higher cost of production, with the Pew Research Center finding that 98\% of content seen on TikTok is produced by just 25\% of content creators \citep{Bestvater2024}.

Using the 2024 U.S.\ presidential election and 239{,}526 images from Instagram, we examine how people utilize generative AI and digital media. We provide three key contributions. First, we find that meme format, not generative AI, is critical to political information sharing online, and is associated with an increase in engagement. Moreover, while synthetic content does not increase engagement, combining generative AI with memes does. Second, we find ideological divergences in generative AI use. Democrat-leaning accounts use AI for in-group support whereas Republican-leaning accounts use it for out-group attacks. Non-traditional, left-leaning outlets are the primary creators of political memes, with an emphasis on different topics largely following issue ownership. Third, we define the phenomenon of \emph{generative memesis}, where memes are no longer created and transmitted from person to person, but mediated by generative AI. More precisely, creation and curation are now distinct stages in the production of spreadable media \citep{Jenkins2013} and serve as a new socio-technical layer within visual communication systems. We discuss the implications of generative AI as a means of production that aids civic participation, and compare how novel technologies in past and present shift visual civic participation.

\section{Literature Review}

As LLMs are a fairly recent phenomenon, and the 2024 United States elections being the first to include them, there is a dearth of prior literature that studies generative AI content during critical political events. However, two adjacent areas of research form the natural trajectory in visuals and political campaigning: political posters and memes. We offer a brief sketch of this history of visual politics, the novel technologies that enabled them, and parallels and comparisons to generative AI today.

\subsection{The First Wave of Visual Politics: Analog Production and Dissemination}
Political posters and cartoons have long been modes of campaigning as means to mobilize and acquire or challenge political power. They are also intimately tied to the history of press technology. Martin Luther's "95 theses," printed in broadsheets, served as an early version of political posters, challenged the Church's authority, and spurred the Reformation \citep{Marshall2017}. During World War I and II, posters were created to recruit soldiers, including the iconic Uncle Sam's "I Want You" image. During the 1960s and 70s, the intention of these posters turned from outward toward domestic issues, such as the Civil Rights Movement and protest against the Vietnam War \citep{Young2016}. Alongside these movements and counter-movements, posters provided the means to attack and counter-attack, utilizing not just satire but absurdism by portraying situations that defy realism. Political posters employ hyperbole to simplify complex political messages and engage audiences emotionally \citep{ElRefaie2003}. Hyperboles help create an emotional appeal, sometimes leading audiences to interpret exaggerated claims literally. This approach is common when trying to evoke strong emotional responses or create a sense of urgency in political communication \citep{Brader2005,Boukes2015}.

As the internet sparked a transition from the analog to digital, fundamental changes occurred to the creation, distribution, and consumption of political images. Producers now encapsulated anyone with an internet connection. The growth of social media served as intermediaries to share these images with consumers of information, lowering distribution costs to a fraction of their physical counterparts. Most importantly, the reciprocal relation between producers and consumers produced a new format: the internet meme.

\subsection{The Second Wave: Memes and Political Participation}
First introduced by Richard Dawkins in \emph{The Selfish Gene} (1976), a meme refers to a concept, behavior, or piece of information that spreads from person to person, analogizing them to genes in how they replicate, mutate, and evolve. He derived "meme" from the Greek word \emph{mimema}, meaning "that which is imitated," blending it with the English word "gene" to emphasize the spread and mutation of ideas. Melodies, catchphrases, and fashion were all things that slowly evolved under competitive pressure.

With the rise of the internet, the term "meme" has evolved to refer specifically to images, videos, or phrases humorously modified or shared across social media platforms \citep{Shifman2013b,WigginsBowers2015,Blackmore2000}. This coincided with the field of memetics, or "the theoretical and empirical science that studies the replication, spread and evolution of memes," which emerged as an active program of research \citep{HeylighenChielens2009}. Since the burgeoning of the field, criticisms have been raised: the definition is somewhat ambiguous, and some researchers remain skeptical of their ability to describe complex human behavior. In defense, \citet{Shifman2013b} contends that the value of memes can be delineated through content, form, and stance. Content refers to the "ideas and ideologies" conveyed, form refers to the physical incarnation such as visual or audio, and stance refers to a meme's "own communication," or the particular frame that it embodies across person to person. These frames are often what provide memes with their intertextual sense of humor.

As social media platforms have become increasingly visual—with TikTok, Instagram, and Snapchat eroding the market share of traditional platforms like Facebook and Twitter—memes have emerged as one of the most suitable formats for political and civic communication. For instance, visual memes targeting Justin Trudeau in the 2019 Canadian general election heavily featured his blackface images and climate change stances, though experimental results yielded marginal effects \citep{VachonChabot2021}. The 2017 UK General Election, commonly referred to as the "youthquake," featured unprecedented turnout amongst younger voters; political scientists turned to the abundance of memes as a possible explanation \citep{McLoughlinSouthern2021}. Memes were also found to be effective persuasive tools during the COVID-19 pandemic \citep{Wasike2022}, and considered for domestic use in influence operations \citep{Zakem2018}. Experimentally, however, their effects were reported to be minimal even across longitudinal experiments, though there is evidence of backlash among those with strong party identification \citep{Galipeau2023}.

The 2024 United States presidential election featured many such instances of memes and spreadable media. In Kamala Harris's 2024 campaign, followers showed support through memes like "coconut tree," taken from one of Harris's speeches, and "brat summer," a trend characterizing a youthful attitude, to enhance her appeal \citep{Murray2024,Li2024Atlantic,Roose2024,Lee2024TIME}. Harris's official campaign account, Kamala HQ, embraced memes as part of her marketing. Similarly, Donald Trump's campaign courted influencers to increase their reach online \citep{Kurtzleben2024}.

Political campaigning strategies can be further differentiated into positive and negative intentions. Positive campaigning typically focuses on presenting a candidate's strengths, policy proposals, accomplishments, and vision for the future. Negative campaigning, in turn, focuses on an opponent candidate's weaknesses, record, character, or policy positions \citep{Baumeister2001,HarringtonHess1996,LauSigelmanRovner2007,SkaperdasGrofman1995}. These strategies often manifest through the use of memes. \citet{PetersAllan2022} introduce the concept of \emph{memetic weaponization}, referring to the deliberate creation and distribution of memes to influence political, cultural, or ideological conflicts, often as part of information warfare. For instance, memes were a mode of information distortion employed by Russian state actors during the 2020 elections \citep{Posard2021} and during positive campaigning in the U.K.'s YouthQuake.

\subsection{Investigating the Third Wave: Generative Artificial Intelligence}

One weakness of this definition of weaponization lies in that it does not differentiate between creation and distribution \citep{Ferrara2020}. Generative AI encapsulates the latter. At its core, generative AI refers to machine learning techniques that produce media artifacts—such as text, images, audio, and video—by learning underlying patterns from training, representing them in high dimensional space, then sampling from them probabilistically \citep{Goodfellow2014,KingmaWelling2014,Vaswani2017}. Crucially, the rapid development of these technologies allow everyday users to produce digital art at high fidelity with low effort. The distinction from automated diffusion to automated creation distinguishes the third wave of political campaigning. 

Traditionally, memes are defined as being transmitted from one person to another, using existing images or frames with minimal editing. Websites existed purposefully to typologize these memes \citep{Pettis2022}, where base frames are often "found" from comics, cartoons, GIFs, or screenshots from popular media, then edited to reflect analogous situations. Generative AI fundamentally disrupts this process by customizing these base frames. Memes are no longer copied person-to-person but mediated and customized through generative AI. In lieu of negative campaigning, "generative memetic weaponization" describes this use for nefarious purposes such as information warfare or mudslinging, and is now available to not just political artists, but everyday users. As such, it is important to note that memes and AI-generated content are orthogonal definitions. One is a format while generative AI is a means of production.

Recent experimental work has begun to examine the 
demand side reactions to AI in campaigns. \citet{JungherrRauchfleischWuttke2025} 
propose a typology distinguishing campaign operations, voter outreach, 
and deception, finding that while the public disapproves of AI use in 
campaigns, deceptive uses are especially norm-violating 
but do not reduce party favorability. Our study complements this 
demand-side evidence by characterizing how AI-generated content is actually produced and circulated.

\subsection{Social Media, AI, and the Information Funnel}

Whereas the previous sections discuss the content, the forums through which this content travels matter. Social media plays a crucial role in shaping contemporary political information environments. Compared to previous technologies, such as television, social media provide audience members an enormous amount of influence over the information that diffuses to others \citep{Neblo2018,Rathje2021,chang2025liberals}. The social media environment is often described as a funnel. The supply or inventory denotes the content made by producers. Content then gains exposure through social networks and the feed-ranking algorithm. The audience then provides engagement with actions such as likes, comments, and shares. Shares provide further dissemination of content to one's immediate social network.

The history of social media and politics can also be told in three waves. After the Arab Spring, social media was hailed as a democratizing force that allowed citizens to leverage the free internet to facilitate protest and civic engagement \citep{Bruns2014}. In the aftermath of the 2016 U.S.\ presidential election and Brexit, social media came under scrutiny due to official campaigns run by foreign actors, such as coordinated campaigns run by Russia's Internet Research Agency \citep{Lukito2020}. Social media has also been scrutinized as sites of information manipulation, such as state-controlled social bots, particularly during elections \citep{Ferrara2020,BessiFerrara2016}. Crucially, the Cambridge Analytica scandal revealed personal data from millions of Facebook users were used without consent, which enabled microtargeting \citep{Hinds2020}.

In 2020, Meta and independent academic researchers embarked on a large-scale collaboration to audit the effects of the algorithm. Most of the papers reported null findings, where there was no evidence of polarization or echo chambers \citep{Nyhan2023,Guess2023,GonzalezBailon2023}. However, \citet{Munger2020} also points out that the algorithm in 2016 is not the same as that in 2020. What these studies do show is a greater emphasis on understanding the algorithm as a mediator of the information funnel. Whereas social media platforms used to be based on social networks, visual-based platforms like Instagram and TikTok have become increasingly algorithm driven.

\subsection{Research Questions}

In this paper, we examine the visual content disseminated during the 2024 United States election. We build upon prior research that investigates the role of social media on electoral public opinion, campaigning, and political engagement \citep{AnsteadOLoughlin2015,VergeerHermans2013,GilDeZuniga2012}, and combine it with studies on deepfakes and generative AI \citep{Qian2023}.

We take particular interest in articulating how memes and generative AI mediate political information. Memes appeal to users who have a strong affinity for absurdist humor \citep{Partlow2021} and coincide with the growth of "unhinged marketing," featuring "unusual imagery, unexpected storytelling, and ridiculous takes" \citep{Noe2023,Alfred2023}. Generative AI provides means for everyday users to achieve absurdism and hyperbole by combining elements. Together, these characterize the phenomenon of \emph{generative memesis}: defined as the use of generative AI to create spreadable media, rather than editing existing frames. More generally, we also build on existing typologies of visual communication, specifically for static images building on existing agendas \citep{PengLuShen2023}. Aesthetic features include the format (memes), production means (generative AI), and content (the existence of people, their facial expressions, and the types of people like politicians). We also introduce visual themes, which are made possible by object detection in large language models (LLMs). Our research questions can be found below:

\begin{itemize}
    \item \textbf{RQ1 (Characteristics):} What examples of AI-generated content, memes, and their combination can be found?
    \item \textbf{RQ2 (Engagement):} What visual elements and formats are associated with the greatest levels of engagement? 
    \item \textbf{RQ3 (Differential Use):} What elements characterize (synthetic) human faces during the 2024 U.S.\ presidential election?
    \item \textbf{RQ4 (Opinion Leadership):} Who were the primary opinion leaders during the 2024 U.S.\ election and how did their use of generative AI diverge? 
\end{itemize}

We define opinion leaders as the most prominent content-producing accounts, operationalized as the top 50 accounts ranked by at least one of three metrics: total post volume, total likes, or highest average likes per post. This composite criterion captures both high-volume suppliers and high-engagement influencers, rather than privileging either dimension alone.

One observation is that AI now occupies every stage of the information funnel: from occupying the actors as social bots, to the design of feed-ranked algorithms, and now finally the process of production. These research questions provide a comprehensive view of not just the content of synthetic and non-synthetic content, but also the supply and demand of this content, as mediated by artificial intelligence \citep{Munger2020,ZhaChang2025GenderInequalities}.

\section{Methods}

\subsection{Data Scraping and Curation}
To collect Instagram data, we first generated a list of keywords and hashtags relating to the 2024 U.S.\ presidential election. We then used Meta's tool, CrowdTangle, to query for Instagram posts containing one of these keywords. The CrowdTangle API retrieved data including post date, description, URL, favorite (like) count, and comment count, as well as data of the posting account such as account handle and subscriber count. Our dataset is from April 5, 2024 (six months before the election) to August 9, 2024. CrowdTangle shut down on August 14th, ending our data collection from this pipeline. We then downloaded the image and video content of the curated Instagram posts. This yielded 239{,}526 images.

Our keyword list was generated initially from popular election-related hashtags and the names of key election figures. We then used snowball sampling to identify other high-frequency hashtags from the posts within our dataset and expanded our list to include them. The full list can be found in Appendix~A. As 2024 is the largest election year in history, we further subset our data to direct mentions of politicians and the United States election. This yielded a total of 285{,}890 images. To identify the opinion leaders, we filtered for the top 50 content-producing accounts for at least one of the following metrics: most posts, most likes overall, and highest average likes. Rather than measuring only the total volume of supply, top content creators depend on their influence, measured by engagement and reach of the content. This union-based criterion yielded 139 unique accounts.

\subsection{Visual Element Identification}
To identify visual elements, we utilized the OpenAI API to access the GPT-4o-mini large language model, which features robust image detection capabilities. We include our full prompt in Appendix~B. We employed a zero-shot labeling approach using GPT. Zero-shot labeling has seen significant recent use in social scientific research, as it leverages pre-trained models to classify text at a cheap cost~\citep{Munker2025, Neuman2023}. This extends to visual content especially in extracting thematic objects. Each image was downsampled and encoded before being sent via the API with a prompt requesting specific classifications, such as probabilities of protest, presence of national flags, war, the presence of key political figures, and other thematic content. We also use zero-shot labeling to determine the probability of whether a post  adopts a meme format. The thematic labels (e.g., protest, war, immigration, economy) were selected a priori based on the dominant policy issues identified in the political communication literature on the 2024 election cycle, supplemented by salient topics from preliminary topic modeling. The GPT-annotator was provided a definition of a meme\citep{Shifman2013b,WigginsBowers2015}. The GPT model provided probability estimates for each category without prior training on the dataset, and the results were saved in batch files for further analysis. To label whether an image was synthetically generated, we used Provenance, a generative AI detection tool developed by OpenAI. Manual analysis was conducted to verify the presence of topics in posts labeled by the automated labeling system. We also manually assessed whether posts were clearly AI-generated to validate the performance of Provenance. To assess labeling quality, a random sample of 500 images was independently coded by two research assistants. Across all topics and synthetic content, manual coding indicated that the automated labels achieved an accuracy of approximately 85–92\%.

Because AI-generated content and meme format are orthogonal dimensions, with one denoting means of production, the other denoting communicative format, we annotate each independently and then cross-classify. An image is labeled ``AI+meme'' if it is both flagged as synthetic by Provenance \emph{and} classified as adopting meme format by the GPT labeler. 

To rank opinion leaders, we weight each account's post count by its average view count. Formally, for account $j$ with $n_j$ posts and average views $\bar{v}_j$, the weighted score is $S_j = n_j \cdot \bar{v}_j$. Accounts are then ranked by $S_j$ to identify the top content producers.

\subsection{Labeling Affect}
Many of these images include human faces. While this can also be labeled using zero-shot approaches, a more accurate alternative exists grounded in psychology. For text, sentiment analysis became a well-established solution for scalable analysis. However, this fails with image-based media. Grounded in psychology, facial emotion recognition lies in the analysis of facial action units (AUs), which are specific muscle movements defined within the Facial Action Coding System (FACS) introduced by \citet{EkmanFriesen1971}. For instance, the combination of AU6 (cheek raiser) and AU12 (lip corner puller) corresponds to a genuine smile. To measure this, we utilized Py-Feat, an open-source tool for working with facial expression data \citep{Cheong2023} that has been extended in recent years to large-scale social media analysis \citep{WeiNohChang2025}.

\section{Results}

\subsection{Examples of Synthetic Content and Memes}
\begin{figure}[!htb]
  \centering
   \includegraphics[width=1.0\linewidth]{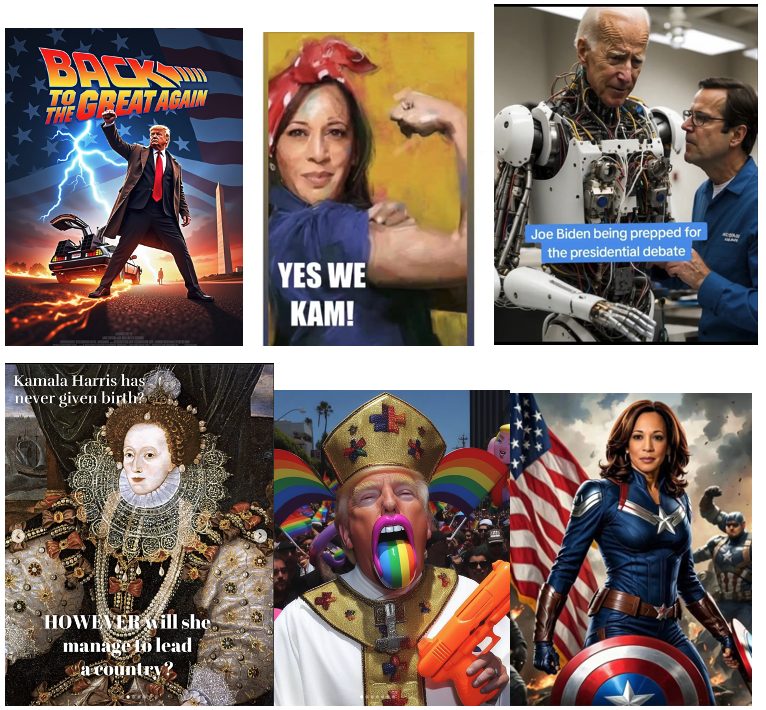}
  \caption{Examples of generative images and memes during the 2024 U.S.\ presidential election on Instagram.}
  \label{fig:examples}
\end{figure}

We first provide some examples of AI-generated memes along the dimensions of positive and negative campaigning. Figure~\ref{fig:examples} shows examples of viral visual content. 

\begin{figure*}[h]
  \centering
\includegraphics[width=0.75\linewidth]{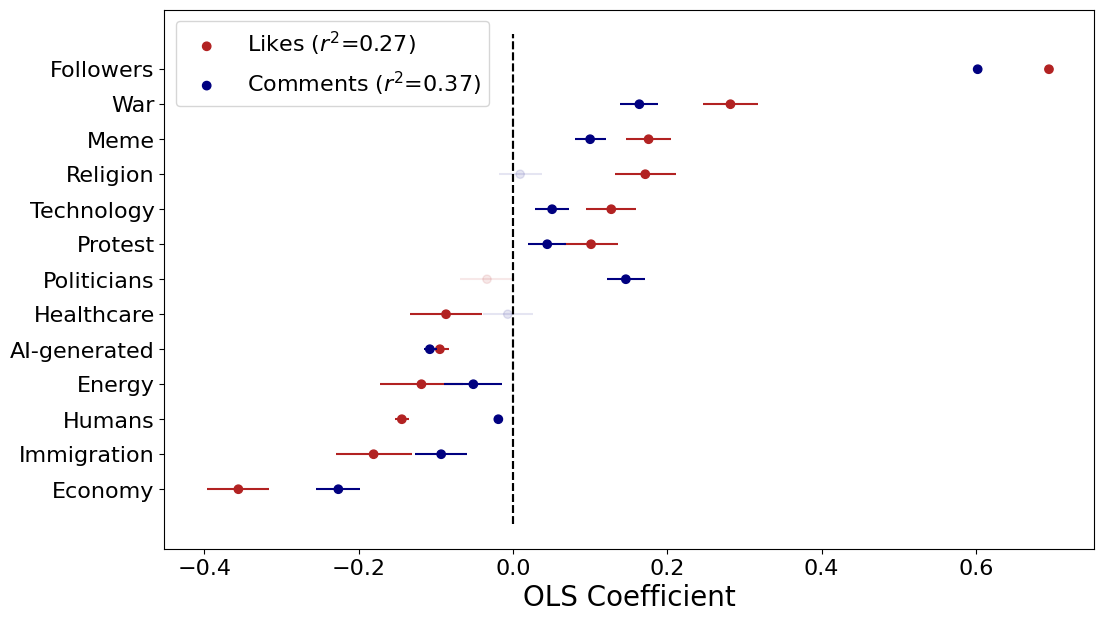}
  \caption{Ordinary Least Squares regression of log Likes (red) and log Comments (blue). Independent variables include key themes in the 2024 U.S.\ presidential election.}
  \label{fig:forest}
\end{figure*}

Figure~\ref{fig:examples}a shows former President Trump in a movie theater poster, a play on the "Back to the Future" franchise. This was shared by a supporter as part of his campaign, a play on Donald Trump's slogan "Make America Great Again." Figure~\ref{fig:examples}b is Harris's face generated over the iconic women's rights poster, with the slogan "Yes, we can." These two examples of generative memetic support combine AI with memes for positive campaigning by accentuating aspects of the candidate. On the other hand, generative memes can be used to attack candidates as well. Figure~\ref{fig:examples}c shows Joe Biden's head on a cyborg with a caption of him "being prepared for the presidential debate." Seen earlier in the campaign trail, this is a case of negative campaigning used to attack Biden's age, and therefore his electability as a candidate.

Next, we consider cases of traditional memes and simple Photoshop. Figure~\ref{fig:examples}d overlays the text of "Kamala Harris has never given birth?" on a portrait of Elizabeth I, from the House of Tudor, also known as the Virgin Queen. Unlike the other examples, this is a traditional meme that overlays text over existing images. Figure~\ref{fig:examples}e is Trump in a religious robe, with a rainbow tongue and orange gun, a reference to policy positions he frequently mentions, as a form of satire. However, in the absence of text, it is unclear what position the message and source take on. Lastly, Figure~\ref{fig:examples}f shows Harris in the uniform of Captain America. The soldier in the back is a case of mistakes made by generative AI, where an arm is growing out of the soldier's head.

In sum, the media environment on Instagram contained a mixture of normal memes, AI-generated content, and memes with AI content. This answers \textbf{RQ1}. Next, we evaluate the extent each of these have on user engagement. 

\subsection{Memes, not AI-Generated Content, Produce Greater Virality and Engagement}

In this study, AI-generated content and whether something is a meme is separately annotated, then cross-sectioned. A post with binary labels of 1 for both categories would be an AI meme. Within all memes, 85\% are traditional memes and 15\% are AI-memes. Memes account for only 10\% of total posts in this election dataset, with the remaining 90\% including photographs, infographics, screenshots, and other visual formats. 

Figure~\ref{fig:forest} shows the forest plot of regression coefficients in estimating Likes (red) and Comments (blue) at the post level. The x-axis shows the coefficients under OLS regression. The y-axis shows independent variables, which include the number of followers as a control, major themes identified in the 2024 U.S.\ presidential election campaign (e.g., immigration and the economy), and the format (e.g., meme and AI-generated content).

For both dependent variables, the number of followers—as expected—most strongly predicts the virality of an image. For Likes, topics such as war, religion, technology, and protest generate the most engagement, while topics like healthcare, energy, immigration, and the economy negatively predict virality. The prevalence of war is likely due to the divisive conflict between Israel and Palestine; religion is thus also positive due to its overlap with the conflict. 
Interestingly, while the inclusion of politicians increases comments—a more active form of engagement—it does not increase the number of likes (the coefficient is slightly negative though not significant).

Crucially, posts that are memes are strongly associated with increases in virality and engagement, while AI-generated content is less viral relative to the universe of election-related content. Although generative AI has garnered more press coverage due to its novelty, the traditional format of a meme—a combination of visuals and text—may be better suited for political communication due to conveying verbal information.

\begin{figure}[!htb]
  \centering
\includegraphics[width=\linewidth]{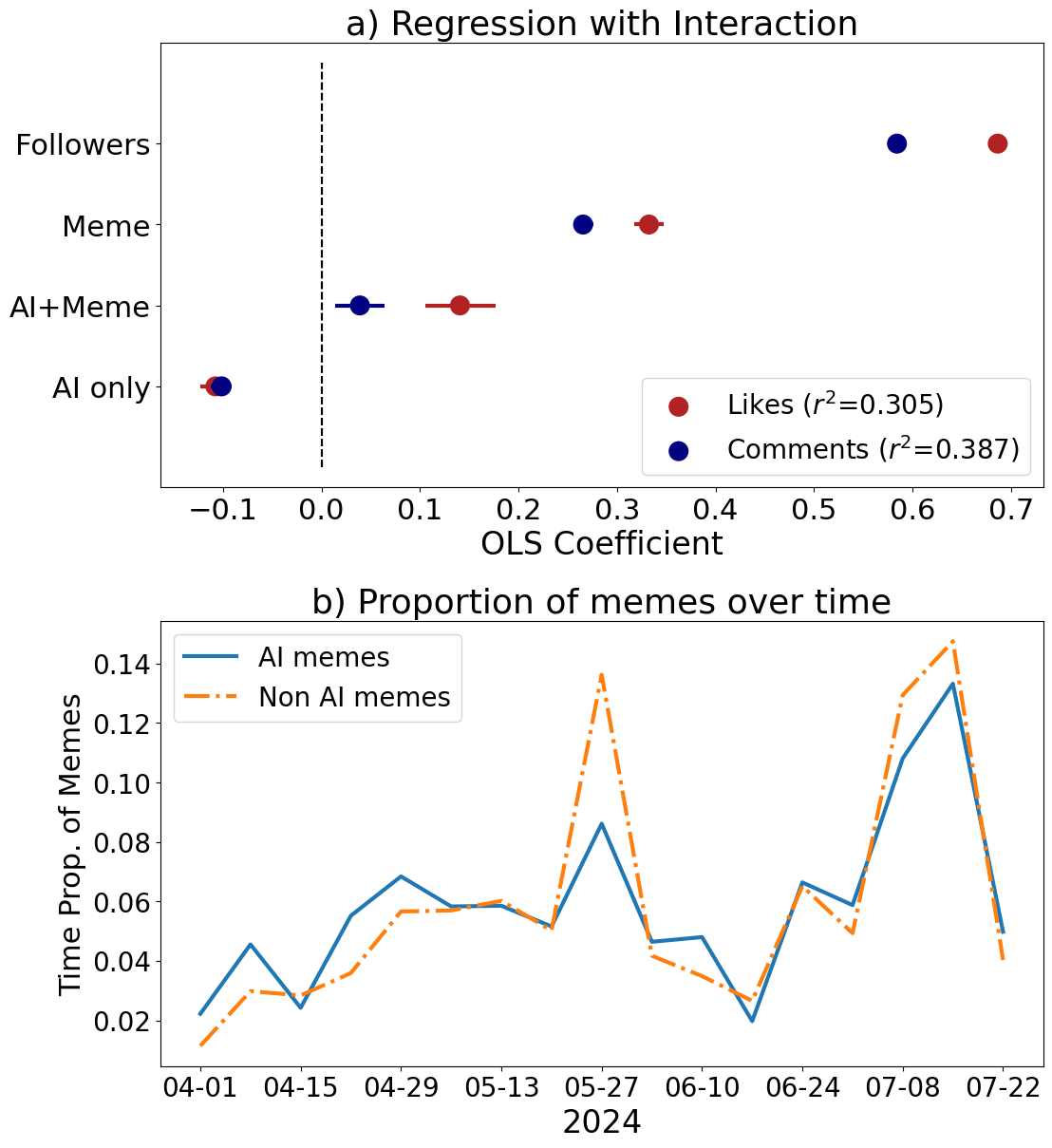}
  \caption{(a) Regression on Likes and Comments with followers, format features, and AI $\times$ meme interaction. The positive interaction term indicates a synergistic effect when AI-generation is coupled with memes. (b) Time series of AI memes versus normal memes.}
  \label{fig:interaction}
\end{figure}

When cross-sectioning on AI-generation status and memes, we find that 15.3\% of memes are synthetic, whereas 84.7\% are non-synthetic. Figure~\ref{fig:interaction}a shows the regression on Likes (red) and Comments (blue) using followers, meme format, whether it was synthetic or not, and the interaction between the latter two variables. Similar to Figure~\ref{fig:forest}, we find a decrease in virality with pure synthetic images, and an increase in the meme format.
However, the positive interaction term indicates that when posts are synthetic memes, we observe a synergistic boost in both Likes and Comments, though this is more pronounced in Likes. 

To further differentiate between normal and synthetic memes, we consider a time series of the two. Figure~\ref{fig:interaction}b shows the time series of synthetic memes (blue) versus normal memes (orange). The activity between AI-generated memes and non-AI-generated memes is largely correlated, peaking at key events such as the assassination attempt on Trump (07/14/2024), Trump's conviction (05/31/2024), Harris announcing her replacing Biden (07/21/2024), and the debate between Biden and Trump (06/28/2024). Of note, the largest gap between AI-generated memes versus non-AI-generated memes arises regarding Trump's conviction (14\% non-AI-generated versus 8\% AI-generated). This indicates underlying events may moderate the extent to which AI-generated content needs to be supplied. This answers \textbf{RQ2}.

\subsection{Selective AI Use by Affect and Ideology}

\begin{figure}[!htb]
  \centering
  \includegraphics[width=\linewidth]{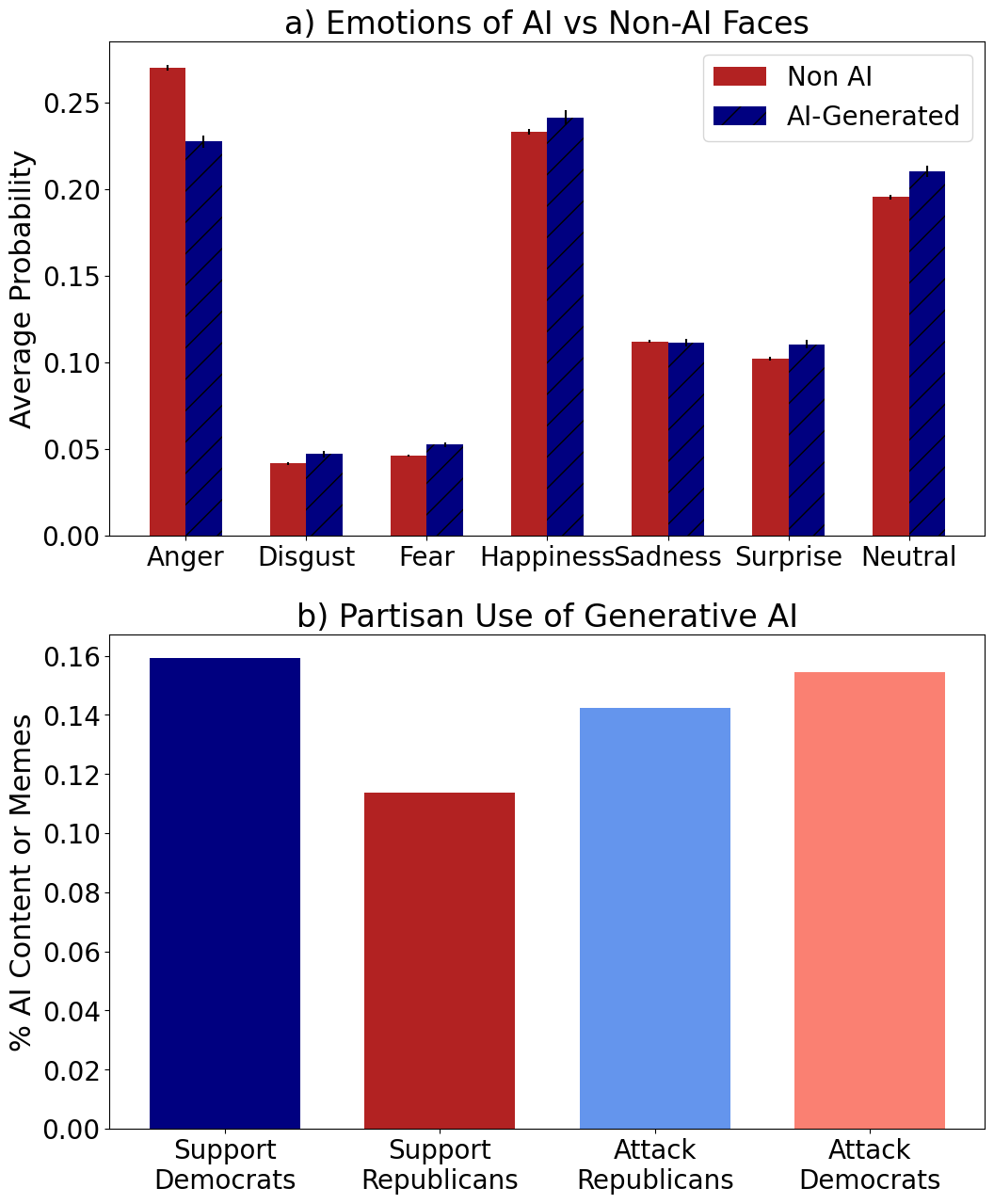}
  \caption{(a) Emotions of generated faces based on psychology metrics. (b) AI-generated and meme content by type of group-based messaging.}
  \label{fig:affect}
\end{figure}

We can also consider if there are differences in how synthetic humans are framed compared to real humans. Drawing from tools in the psychology literature, we employ facial analysis algorithms that quantify human expressions based on the activity of facial muscles \citep{Cheong2023}. Figure~\ref{fig:affect}a shows the emotions demonstrated by AI-generated faces versus non-AI-generated faces. Results show that non-AI-generated faces are around 5\% angrier than AI-generated faces, whereas AI-generated faces are slightly more neutral and happy.

This may be a result of the algorithms themselves or curation from the content creators. 
Using zero-shot labeling, we also label the captions of the figures as either supporting or attacking Democrats or Republicans (92\% accuracy). Figure~\ref{fig:affect}b shows the proportion of AI-generated memes for each message group. We find posts supporting Democrats (Democrat-leaning in-group favoritism) have the highest number of synthetic images, followed by ones used to attack Democrats. The presence of AI-generated images and memes is lowest among posts supporting Republicans. Two implications thus emerge: first, Democrat-leaning supporters tend to use generative AI more; second, there is an asymmetry in how Democrat-leaning supporters use generative AI compared to Republican-leaning supporters---Democrat-leaning accounts use it more for supporting their party (in-group favoritism), whereas Republican-leaning accounts use it more for attacking Democrats (out-group animosity). This answers \textbf{RQ3}.

\begin{figure*}[!htb]
  \centering
\includegraphics[width=0.8\linewidth]{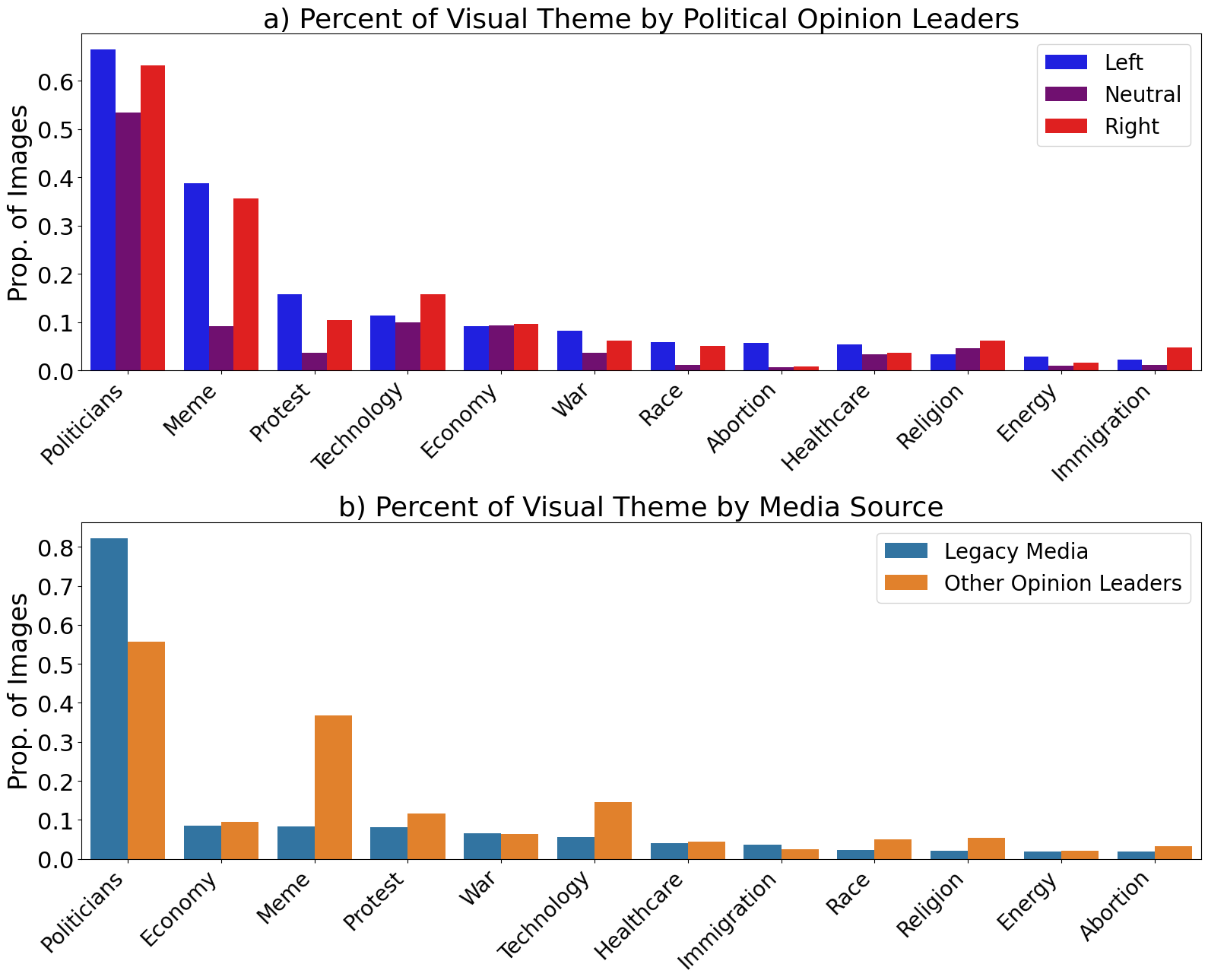}
  \caption{Sources of visual themes such as memes, policy issues, and politician mentions. (a) By ideology. (b) By media type.}
  \label{fig:leaders}
\end{figure*}

\subsection{Opinion Leadership on Instagram: Leaders Utilize Twice the Number of Memes}

Beyond the content of memes and AI-generated content, we also analyze the source. To do so, we sorted the top content creators by the number of their posts weighted by their views (see formula in Section~3.2). The top 139 content producers were then manually labeled based on their media bias, referencing AllSides and Media Bias/Fact Check \citep{Huszar2022}. Appendix~C shows the full list of these opinion leaders, which includes a mix of legacy media accounts and non-legacy media accounts. Among these 139 accounts, memes constituted 27.6\% of their content and AI-generated images constituted approximately 8.4\%, compared to 10\% and 5.2\% respectively across all users in the dataset. For example, The Shade Room (one of the top accounts) is an online publication that specializes in celebrity gossip specifically within the Black community. While entertainment-based in content, they were one of the top content producers during the Floyd protests in 2020, which builds on the theory that non-institutional accounts are dominant players in shaping the visual information ecosystem \citep{ChangRichardsonFerrara2022}. The highly developed audiences of entertainment accounts serve as a conduit for political information.

Figure~\ref{fig:leaders}a shows the ideological leaning of the source and its intersection with themes. Overall, left-leaning sources include more politicians, memes, and imagery about protest and notably abortion, for which there is little mention from neutral and right-wing sources. On the other hand, right-leaning content creators feature more visual references to technology, religion, and immigration. Overall, these findings align with the issue of ownership by the different parties. Issue ownership refers to the perception that a particular political party or group is better at handling or addressing a specific issue compared to others \citep{Peitrock1996}. It stems from a party's historical association with that issue, its policy expertise, or its success in addressing it in the past. This concept plays a significant role in elections, as parties tend to focus on the issues they ``own'' to attract voters and frame the political agenda in their favor.

Figure~\ref{fig:leaders}b shows legacy media focuses on politicians almost 25\% more, whereas non-legacy media opinion leaders utilize many more memes and AI-generated content. Table~\ref{tab:leaders} further summarizes the proportion of likes by the top opinion leaders. Notably, while opinion leader references to politicians are largely the same as overall users, the use of memes doubles (from 13.7\% to 27.6\%). We also observe a much greater value for mean Likes than median Likes, which is consistent with power law distributions—in cases where the top producers command disproportionate attention (i.e., number of likes and comments) there will be heavy tails. In short, opinion leaders on Instagram tend to disproportionately use the meme format for their political communication.

\begin{table*}[t]
\centering
\begin{tabular}{@{}lrrrrr@{}}
\toprule
Theme & \% Images & Mean Followers & Mean Likes & Median Likes & Mean Comments \\
\midrule
Politicians & 57.3 & 2{,}706{,}619 & 16{,}668 & 1{,}425 & 957 \\
Meme        & 27.6 & 1{,}267{,}804 & 11{,}653 & 426   & 469 \\
Technology  & 11.8 & 2{,}099{,}239 & 15{,}318 & 661   & 737 \\
Economy     & 10.3 & 1{,}424{,}880 & 10{,}147 & 330   & 588 \\
Protest     & 10.1 & 1{,}992{,}055 & 18{,}089 & 1{,}480 & 806 \\
War         & 6.0  & 1{,}646{,}949 & 15{,}899 & 1{,}095 & 569 \\
Religion    & 4.4  &   914{,}111 &  8{,}909 &   80 & 316 \\
Healthcare  & 4.1  & 2{,}102{,}782 & 11{,}028 & 1{,}582 & 565 \\
Race        & 3.8  & 1{,}898{,}110 & 13{,}811 & 1{,}532 & 926 \\
Immigration & 2.6  & 1{,}615{,}426 & 10{,}974 & 375   & 698 \\
Abortion    & 2.6  & 1{,}656{,}168 &  8{,}447 & 4{,}097 & 596 \\
Energy      & 1.9  & 1{,}173{,}272 & 10{,}378 &   760 & 344 \\
\bottomrule
\end{tabular}\caption{Percentage of visual themes and summary statistics for top opinion leaders.}
\label{tab:leaders}
\end{table*}

In sum, the analysis of opinion leaders shows how a small set of highly visible accounts disproportionately shape visual political information. While legacy outlets emphasize politicians, non-legacy and entertainment-based accounts lean heavily on memes and AI-generated imagery. Issue ownership and media type intersect with format choices, producing a heavy-tailed distribution of engagement. Together, these answer \textbf{RQ4}.

\section{Discussion}

Prior to the 2024 U.S.\ presidential election, the news media reported concerns about generative AI in information manipulation. PolitiFact reported on "Not-so-innocent 'cat memes'," where AI creations produced content related to the unsubstantiated claim of immigrants eating pets in Springfield, Ohio \citep{Tuquero2024a}. A few months later, the same agency reported on the surprising lack of impact AI had \citep{Tuquero2024b}, and this was documented by multiple media outlets \citep{Times2024,AlJazeera2024,Politifact2024}. These reports run somewhat counter-intuitive to concerns following social bots and automated agents in both the 2016 and 2020 U.S.\ presidential elections \citep{Ferrara2020}. Our results corroborate these findings, with insight as to the reasons why, and meaningfully extend prior literature on memetics \citep{Shifman2013b}, spreadable media \citep{Jenkins2013}, visual political engagement \citep{BrownMourao2021,McLoughlinSouthern2021}, and direct experimental evidence of its limited impact~\citep{JungherrRauchfleischWuttke2025}.

First, our analysis suggests that generative AI's current role is best understood as a tool of production, not dissemination. Whereas previous AI agents—like social bots—functioned within the distribution layers of the information funnel, generative AI shifts influence to the supply layer. This reconfigures communicative agency: rather than obscuring who is sharing what, generative AI calls into question who is speaking. Here, our findings resonate with concerns about source credibility and perceived authenticity \citep{Metzger2003,TomaHancock2012}, which may explain lower engagement with purely synthetic content. 

Second, generative AI bifurcates the supply of information, which is associated with decreased engagement. Whereas the meme format increases engagement between 0.25 to 0.35 standard deviations, AI content actually decreases engagement rates. However, when generative AI is combined with meme format, there is a slight boost in engagement, especially for comments. As such, if we are to view generative AI as bifurcating the supply-side of information into two parts—into production and curation—then those that require more human agency produce a greater level of engagement. Indeed, we find that the vast majority of memes (85\%) in this dataset are in the traditional, non-AI format. These posts yield greater levels of engagement, though our analysis does not preclude the possibility of selection bias (i.e. popular accounts use more generative AI). 

While generative AI enables the mass production of visual political content, our findings underscore that spreadability---as theorized by \citet{Jenkins2013}---remains contingent on human intervention. Specifically, the resonance and circulation of synthetic political memes depend not on generation alone, but on the capacity of users to identify and align content with culturally salient frames and genre conventions. In this sense, human actors serve as cultural intermediaries \citep{Bourdieu1984,Couldry2012}, filtering and elevating content that fits within recognizable political narratives and meme formats. Rather than displacing human agency, generative AI shifts it toward tasks of selection, contextualization, and frame calibration \citep{Entman1993}. Memes do not become spreadable simply because they are algorithmically novel; they become spreadable when they are semantically legible to specific interpretive communities \citep{Shifman2013a}.

Our results show multiple instances of how humans actively curate this content, frequently with some form of psychological bias. Users tend to post synthetic human content with happier expressions, whereas photos are more negative, especially angrier, overall. Compared to studies that find negativity spreads faster and farther on social media \citep{Kramer2014}, human curation of AI-generated images emerges as naturally more positive. What this implies is that generative AI bifurcates the stages of supply—generation and curation are now distinct stages in the production of content. This builds on efforts to delineate divergences in the supply and demand side of harms, particularly cases where visual content may actually lean more positive~\cite{ZhaChang2025GenderInequalities}. Additionally, we identify a partisan asymmetry: Democrat-leaning accounts tend to use AI for positive campaigning (16\% positive versus 14\% negative) whereas Republican-leaning accounts use it for negative campaigning (11\% positive versus 15.5\% negative). Further study utilizing synthetic content can further elucidate how partisans react at a mechanistic level. Lastly, especially on social-network-based platforms like Instagram, non-institutional accounts are the primary disseminators of meme content and effective opinion leaders that serve as interfaces of political information.

There are a few limitations to our study. One limitation of this labeling procedure is that some themes are better expressed textually than visually or require multimodal cues. For instance, while the topic of war is easily identified through the presence of troops or refugees, the topic of race requires clarification from text within the image or via the caption. While our sampling generates high levels of accuracy, especially compared to prior generations of automated labeling, a multimodal analysis deserves further investigation. A second limitation is using Py-feat for labeling emotions on AI-generate faces. Since Pyfeat was only trained on real human faces and uses AUs, it is unclear if Py-feat accurately captures emotions for AI faces. This deserves further investigation in future studies. Another potential limitation is that some some of the accounts in our dataset could be bots. However, compared to Twitter studies where resharing takes on significant diffusion volume, and lack of API integration, the barrier to production is likely higher for Instagram, where content creation requires image or video assets rather than simple text retweets. Another limitation is that we only consider Instagram for this study. Substantively, this is an analysis of one platform for one election. While Instagram is the largest social media platform for static images, considering the AI images on platforms such as TikTok, X, or Facebook may show different dynamics. 

Each platform's affordances (i.e. algorithmic curation, content format constraints, and audience demographics) shape how AI-generated content circulates, and patterns observed on Instagram may not generalize to platforms with different production and distribution norms. Running similar observational and experimental analyses in future elections or political events will aid the generalizability of these findings. And as such, further research that associates AI adoption with supply-side content creation could more concretely characterize causal effects.

On generalizability, it is also important to note that while generative visual AI is primarily a means of production, rather than distribution, continuous evolutions may produce "platformification" of these tools. For instance, Sora has distribution capabilities ~\cite{Sora2025}. In tandem, text-based AI services like ChatGPT and Claude have become default search engines and therefore both generators and distributors of knowledge. 

In sum, this research extends existing work that shows the increasingly important role of memes, entertainment-based opinion leaders, and social media in facilitating sociopolitical information. Generative AI is reshaping the political information environment—not by displacing human agency, but by reconfiguring it. While automated systems now assist in the creation of political visuals, their communicative efficacy remains contingent on human actors who select, frame, and contextualize content for circulation. The notion of \emph{generative memesis} captures this shift: a hybrid media logic in which algorithmic generation and human curation jointly produce spreadable political content. As generative systems become more deeply embedded in media infrastructures, future research should investigate how different communities develop norms around synthetic curation, how platform affordances shape the spreadability of AI-mediated content, and how these dynamics influence not just visibility but persuasion and belief formation. Understanding these evolving configurations of automation and agency will be critical for assessing the democratic implications of generative media.

\section{Acknowledgements}

We thank Adrian Rauchfleisch and Brendan Nyhan for their valuable feedback on this work. We acknowledge OpenAI's provenance tool. We also thank the anonymous reviewers for their helpful comments and suggestions.

\bibliography{aaai2026}

\clearpage

\appendix
\section{Appendix}

\section{List of keywords and hashtags}

Kamala, Harris, kamalaharris, Kamala Harris, kamalaharris2024, Voteharris, Coconuttree, Vance, JD Vance, jdvance, Trumpvance, trumpvance2024, Debate, Presidentialdebate, walz, timwalz, election2024, US Elections, uselections, 2024Elections, 2024 elections, 2024PresidentialElections, Project2025, Project 2025, Biden, presidentbiden, bidenharris, JoeBiden, Joe Biden, joseph biden, biden2024, bidenharris2024, trump, donald trump, donaldtrump, donaldtrump2024, trump2024, presidenttrump, Trumpismypresident, trumpsupporters, trumptrain, republicansoftiktok, conservative, maga, saveamerica2024, makeamericagreatagain, ultramaga, KAG, Republican, republicans, RNC, republican party, GOP, CPAC, letsgobrandon, democratsoftiktok, democratsarehot, thedemocrats, voteblue2024, voteblue, DNC, dnc, democratic party, Democrat, democrats, Third party, third party, Green party, green party, Independent party, RFKJr, RFK Jr., RFK Jr, RFK, rfkjr, rfkjr., rfk jr, rfk jr., Robert F. Kennedy Jr., robert f. kennedy jr., jill stein, jillstein, cornelwest, NikkiHaley, nikkihaley


\section{Prompt for Image Labeling}

\begin{tcolorbox}[breakable,
    title=System prompt,
    colback=gray!10,    
    colframe=gray!70,   
    fonttitle=\bfseries,
    boxrule=0.8pt,      
    arc=2pt,            
    left=6pt,right=6pt,top=6pt,bottom=6pt 
]
"This is an image that I want to upload. Describe what is in it. " +

"Please also provide the probability that the image contains some form of protest (label as Protest:). " +

"Please also provide the probability that the image contains a national flag, and if so, which country/countries (label as Flag:). " +

"Please also provide the probability that the image contains reference to technology (label as Technology:). " +

"Please also provide the probability that the image contains religious themes or religious symbols (label as Religion:). " +

"Please also provide the probability that the image depicts ongoing war such as Israel/Gaza, Russia/Ukraine, or China/Taiwan (label as War:). " +

"Please also provide the probability that the image depicts a political rally and/or politicians (label as Politicians:). " +

"Please also provide the probability that the image's content relates to climate change and/or energy (label as Energy:). " +

"Please also provide the probability that the image's content relates to the economy (label as Economy:). " +

"Please also provide the probability that the image's content relates to immigration (label as Immigration:). " +

"Please also provide the probability that the image's content relates to healthcare (label as Healthcare:). " +

"Please also provide the probability that the image's content relates to abortion (label as Abortion:). " +

"Please also provide the probability that the image's content relates to issues of race (label as Race:). " +

"Please also provide the probability that the image is a meme (label as Meme:). " +

"Please also state whether the image is most likely a real photo, photorealistic AI-generated, or absurd AI-generated. (label as AI:). " +

"Please also state whether the image contains a human (label as Human:). " +

"If it contains a human, who are they, how many are there, what is their race, gender, age, and what emotions are they displaying. " +

"If Joe Biden or Kamala Harris or Donald Trump or J.D. Vance is depicted, please indicate so." +

"If it contains text, what language is it (label as Text:)? " +

"Limit the description to 20 words. Treat each image separately. Make sure labels are consistent. " +

"The following formatting notes are extremely important to follow exactly correctly: " +

"Please give probabilities as percentage likelihood (i.e. 1\% if very unlikely and 99\% if extremely likely). " +

"Please begin the response for an image with a number (ex. 1., 2., etc.) " +

"Please separate the response for each image with '\$'." +

"Please separate the responses to questions with '|'. "
\end{tcolorbox}

\onecolumn

\section{Top Opinion Leaders}

\begin{longtable}{l p{1.5cm} p{6.5cm} c c}
\toprule
\textbf{Account Handle} & \textbf{Ideology} & \textbf{Notes} & \textbf{Relevant} & \textbf{Legacy Media} \\
\midrule
\endfirsthead

\toprule
\textbf{Account Handle} & \textbf{Ideology} & \textbf{Notes} & \textbf{Relevant} & \textbf{Legacy Media} \\
\midrule
\endhead

\bottomrule
\endfoot

sheepleskateboards\_mkultra & Right & 9/11 truther account & Y & N \\
courtneyclift & Left & Political cartoonist & Y & N \\
occupydemocrats & Left & Political organization & Y & N \\
naija\_pr & Left & Non-binary Nigerian social media blogger & Y & N \\
coveringpolitics & Left & Newspaper (Washington Post) & Y & Y \\
spencerdgray & Left & Memes, political subversion, communism? & Y & N \\
hollywoodunlocked & Left & Hollywood figure and city council rep & Y & N \\
washingtonpost & Left & Newspaper (Washington Post) & Y & Y \\
nowthisimpact & Left & Social media news organization & Y & N \\
graftacus & Left & Anti-trump fan page & Y & N \\
americanistmemes & Left & Progressive meme page & Y & N \\
alanncalifornia & Left & Leftist, anti-trump page & Y & N \\
red\_maat & Left & Yoga and wellness company, also Black abolitionist & Y & N \\
theshaderoom & Left & Black social media news & Y & N \\
noticiasestrellatv & Left & Spanish language news media & Y & N \\
msnbc & Left & Newspaper (MSNBC) & Y & Y \\
abcnews & Left & Newspaper (ABC News) & Y & Y \\
calltoactivism & Left & Progressive media personality & Y & N \\
thatsnotrightpolitics & Left & Progressive social media and news page & Y & N \\
aljazeeraenglish & Left & Newspaper (English version of Al Jazeera) & Y & Y \\
kittenwhip & Left & Anti-trump fan page & Y & N \\
abnormalize.being & Left & Leftist, anti-liberal memes and news & Y & N \\
\_stillwerise & Left & Focus on social justice and progressive issues & Y & N \\
nytimes & Left & Newspaper (New York Times) & Y & Y \\
ajplus & Left & Progressive news, promoting human rights & Y & N \\
ckyourprivilege & Left & Antiracism education platform & Y & N \\
landpalestine & Left & Advocacy for Palestinian rights (English) & Y & N \\
wissamgaza & Left & Pulitzer-winning photographer, Gaza issues & Y & N \\
everydaypalestine & Left & Advocacy and news on Palestinian issues & Y & N \\
climatepower & Left & Focus on climate issues, pro–Biden-Harris & Y & N \\
lowkeyonline & Left & Musician posting pro-Palestine content & Y & N \\
cnn & Left & Newspaper (CNN) & Y & Y \\
sainthoax & Left & General news, some memes & Y & N \\
osopepatrisse & Left & Artist and abolitionist & Y & N \\
impact & Left & General social issues content & Y & N \\
khaledbeydoun & Left & Law professor, civil rights and Muslim issues & Y & N \\
environment & Left & Incorporated climate advocacy source & Y & Y \\
poderespretos & Left & Portuguese-language focus on Black rights & Y & N \\
eye.on.palestine2 & Left & Palestinian advocacy & Y & N \\
sapphiracristal & Left & LGBTQ+ activist and drag performer & Y & N \\
michelleobama & Left & Official account of Michelle Obama & Y & N \\
eye.on.palestine & Left & Advocacy and news on Palestinian issues & Y & N \\
sbeih.jpg & Left & Palestine news and advocacy & Y & N \\
aoc & Left & Account of Rep. Alexandria Ocasio-Cortez & Y & N \\
lulaoficial & Left & President Lula of Brazil & Y & N \\
misha & Left & Actor, pro-Harris & Y & N \\
jpeachyliberty & Right & Political themed meme page & Y & N \\
trentsteele\_libertydad & Right & Meme page & Y & N \\
allsidesnow & Neutral & Organization rating media bias & Y & N \\
theinvestingplanet & Neutral & AI, startup, business news & Y & N \\
voanews & Neutral & Newspaper (Voice of America) & Y & Y \\
thehill & Neutral & Newspaper (The Hill) & Y & Y \\
us.elections & Neutral & Polling and election results & Y & N \\
npr & Neutral & Newspaper (NPR) & Y & Y \\
bbcnews & Neutral & Newspaper (BBC) & Y & Y \\
washingtonexaminer & Right & Newspaper (Washington Examiner) & Y & Y \\
redstatetshirts2 & Right & Trump fan page and T-shirt company & Y & N \\
your\_first\_million & Right & Author, avid Trump supporter & Y & N \\
themisfitpatriot & Right & Rumble personality & Y & N \\
freedomusa\_\_119 & Right & Conservative meme page & Y & N \\
realmichaelsolakiewicz & Right & Trump fan page & Y & N \\
newsmax & Right & Newspaper (Newsmax) & Y & Y \\
conservative\_womenofusa & Right & Conservative women youth leader & Y & N \\
lutheran.monarchist.v2 & Right & Evangelical Catholic white supremacist page & Y & N \\
bullmoose\_memes & Right & Conservative meme page & Y & N \\
wearetriumphers & Right & Black Christian pro-Trump personality & Y & N \\
wearebreitbart & Right & Newspaper (Breitbart) & Y & Y \\
foxnews & Right & Newspaper (Fox News) & Y & Y \\
americafirstgop & Right & Pro-Trump and anti-Harris content & Y & N \\
lhdaoficial & Right & Spanish language news and memes & Y & N \\
trump.now & Right & Pro-Trump news and memes & Y & N \\
red.wave1776 & Right & Conservative pro-Trump content & Y & N \\
republicanparty1776 & Right & Republican Party and Trump & Y & N \\
teamtrump & Right & Official Trump campaign account & Y & N \\
officialegyptx & Right & Mental health advocate, pro-Trump & Y & N \\
charliekirk1776 & Right & Political commentator, Turning Point & Y & N \\
realdonaldtrump & Right & Official account of Donald Trump & Y & N \\
the\_typical\_liberal & Right & Aspiring political candidate, pro-Trump & Y & N \\
dc\_draino & Right & Lawyer, pro-Trump content & Y & N \\
ragingpatriots & Right & Conservative meme/news page & Y & N \\
dc.swamp & Right & Conservative meme/news page & Y & N \\
therickwilson & Right & Writer, Lincoln Project member & Y & N \\
ragingamericans & Right & Conservative meme and news page & Y & N \\
bennyjohnson & Right & Conservative political commentator & Y & N \\
robertfkennedyjr & Right & RFK account, anti-vax & Y & N \\
millennial\_republicans & Right & Conservative content for Millennials & Y & N \\
serio.330 & Right & Mexican American rapper, pro-Trump & Y & N \\
theredelephantt & Right & Mostly anti-Harris and some pro-Trump & Y & N \\
republicans.genz & Right & Gen-Z conservative content & Y & N \\
hulkhogan & Right & Pro wrestler, endorsed Trump & Y & N \\
thegeostrata & Left & Indian youth-led think tank & S & N \\
midianinja & Left & Portuguese journalism and activism & S & N \\
the\_hindu & Left & Indian newspaper & S & Y \\
random\_\_predictions & Neutral & Political mapper and prediction creator & S & N \\
bostonculture & Neutral & Culture / art / entertainment / memes & S & N \\
esperancadebate & Neutral & Portuguese news source & S & N \\
nojumper & Neutral & Entertainment and celebrity news & S & N \\
tagesschau & Neutral & German news source & S & N \\
thetrillionairelife & Neutral & Luxury magazine, some Trump news & S & N \\
worldstar & Neutral & HipHop news and viral content. & S & N \\
hoodclips & Neutral & Entertainment and viral clips. & S & N \\
brasilparalelo & Neutral & Brazilian media production company & S & N \\
wealth & Neutral & Business and tech news & S & N \\
memezar & Neutral & Incorporated general meme content. & S & N \\
pubity & Neutral & Incorporated general media content & S & Y \\
realhinakhan & Neutral & Actor, some voting advocacy & S & N \\
complex & Neutral & Incorporated general entertainment and news & S & Y \\
complexpop & Neutral & Incorporated pop culture news & S & Y \\
losherederosdealberdi & Right & Argentinian conservative political page & S & N \\
tejashwipdyadav & Left & Indian politician, leader of the RJD party & N & N \\
rohini\_yadav\_rjd & Left & Indian politician from RJD, daughter of Lalu Prasad Yadav & N & N \\
bigbosz\_style & Neutral & Indian clothing store & N & N \\
swarnam\_varnam & Neutral & Luxury jewelry company & N & N \\
popularmalaysia & Neutral & Trilingual bookstore in Malaysia & N & N \\
torcha & Neutral & Italian social news media page & N & N \\
thetatvaindia & Neutral & Indian general media account & N & N \\
insidehistory & Neutral & General memes & N & N \\
esthetic.cutz\_ & Neutral & South and Southeast Asian song edits & N & N \\
ennam\_pol\_\_vazhkaii & Neutral & Tamil personal page, not politically focused & N & N \\
instantbollywood & Neutral & Bollywood entertainment news. & N & N \\
\_blockaye & Neutral & HipHop, football, celebrity entertainment & N & N \\
dailymusicas & Neutral & Spanish language music content & N & N \\
blackvibe.s & Neutral & Spanish language music content & N & N \\
voguemagazine & Neutral & Fashion and lifestyle magazine & N & N \\
the.meethling & Neutral & Indian actress & N & N \\
ginaindelicada & Neutral & Portuguese (Brazil) memes and news & N & N \\
redbullracing & Neutral & Official account of Red Bull Racing & N & N \\
spotifyindia & Neutral & Official account of Spotify India & N & N \\
hypewhip & Neutral & Car hype culture media & N & N \\
anuuveee & Neutral & Indian Malayalam-speaking actor & N & N \\
siju\_sunny & Neutral & Indian actor & N & N \\
saafboi & Neutral & Indian social media influencer & N & N \\
9gag & Neutral & Entertainment and memes & N & N \\
tntsportsbr & Neutral & Brazilian sports network & N & N \\
oduataj & Neutral & African culture and entertainment & N & N \\
taiwooduala & Neutral & African culture and lifestyle content & N & N \\
afroawardss & Neutral & African entertainment and awards & N & N \\
gossiproomoff & Neutral & French gossip and entertainment news & N & N \\
conexaopoliticabrasil & Right & Brazilian conservative news and political content & N & Y \\

\end{longtable}
\twocolumn

\end{document}